\documentclass[useAMS,usenatbib]{mn2e}

\usepackage{amsmath}
\usepackage{graphicx}
\usepackage{color}

\title[Dependence of the Cosmic Microwave Background Lensing Power Spectrum on the Matter Density]
{Dependence of the Cosmic Microwave Background Lensing Power Spectrumon the Matter Density}
\author[Z. Pan et.al]{Z.~Pan, $^1$\thanks{Email: zhpan@ucdavis.edu}
        L.~Knox,$^1$\thanks{Email: lknox@ucdavis.edu}
   and M.~White,$^2$\thanks{Email: mwhite@berkeley.edu}\\
$^1$Department of Physics, University of California,One Shields Avenue, Davis, CA, USA 95616\\
$^2$Department of Physics and Astronomy, University of California, Berkeley, CA, USA 94720 }

\date{\today}

 \newcommand{\ltsima}{$\; \buildrel < \over \sim \;$}
 \newcommand{\ltsim}{\lower.5ex\hbox{\ltsima}}

\def\be{\begin{equation}}
\def\ee{\end{equation}}
\def\bea{\begin{eqnarray}}
\def\eea{\end{eqnarray}}

\def \nn{\nonumber}
\def \eq{equation}
\def \eqn{eqnarray}

\begin{document}
\label{firstpage}

\maketitle

\begin{abstract}
The anisotropies in the cosmic microwave background (CMB) provide our best
laboratory for testing models of the formation and evolution of large-scale
structure.
The rich features in the cosmic microwave background anisotropy spectrum,
in combination with highly precise observations and theoretical predictions,
also allow us to simultaneously constrain a number of cosmological parameters.
As observations have progressed, measurements at smaller angular scales have
provided increasing leverage.  These smaller angular scales provide sensitive
measures of the matter density through the effect of gravitational lensing.
In this work we provide an analytic explanation of the manner in which the
lensing of CMB anisotropies depends on the matter density, finding that the
dominant effect comes from the shape of the matter power spectrum set by the decay
of small-scale potentials between horizon crossing and matter-radiation
equality.
\end{abstract}


\bigskip\bigskip

\section{Introduction}

Observations of the cosmic microwave background (CMB) temperature power
spectrum have provided us with highly precise determinations of
cosmological parameters \citep{PlanckCollaborationXVI.2013, Hinshaw2013}.  
These determinations are widely used to
aid in the interpretation of other cosmological observables and to
search for possible failures of the standard cosmological model that
may point us toward new physics.  Highly precise inferences are possible
due to the way the rich features in the power spectrum depend on the
underlying model parameters, the ability to compute the spectrum with
high accuracy \citep{Seljak2003} 
and recent high-precision measurements of the anisotropy \citep{PlanckCollaborationI.2013}.
Given the widespread use of these parameter inferences, it is important that
we acquire a physical understanding of how the parameter constraints arise.
Fortunately, the anisotropies arise from relatively simple physical processes
so such an understanding is possible \citep{Hu2002}.

As emphasized by \citet{2014ApJ...782...74H}, gravitational lensing of
the CMB temperature anisotropy power spectrum allows for constraints
on the matter density from measurements of the small-scale (high $\ell$)
portion of the spectrum.
Such precise measurements are now becoming available thanks to the Planck
\citep{Planck},
South Pole Telescope \citep[SPT;][]{SPT} and
Atacama Cosmology Telescope \citep[ACT;][]{ACT}
collaborations.
However, a physical understanding of the dependence of the CMB lensing power
spectrum on the matter density has so far been missing.
In this article we provide an understanding of this dependence, within the
context of the $\Lambda$CDM model.
We will find that it is the decay of the potentials after horizon crossing
due to the presence of radiation which drives the majority of the dependence
and give a simple scaling relation which captures this dependence.

In Section \ref{sec:lowell} we review the origin of the sensitivity of
large-angular scale CMB temperature measurements to the physical matter
density, $\omega_m\equiv\Omega_m h^2$.
In Section \ref{sec:lensCl} we introduce the lensing potential power
spectrum and the basic equations we will use.  With those preliminaries,
we go on in Section \ref{sec:matter} to demonstrate qualitatively and
quantitatively the origin of the sensitivity of the CMB lensing potential
power spectrum to $\omega_m$.
We present our major conclusions in Section \ref{sec:discussion}.

\section{The spectrum below $\ell=1000$ and $\omega_m$}
\label{sec:lowell}

The rich structure in the first few acoustic peaks allows us to simultaneously
constrain a number of cosmological parameters, including the matter density.
Here we briefly review the physics behind these constraints
\citep[see e.g.][for textbook discussions]{Peacock99,LidLyt00,Dodelson03}.

The peaks in the CMB spectrum arise from gravity-driven acoustic oscillations
in the primordial, baryon-photon plasma before recombination.
For nearly scale-invariant, adiabatic fluctuations the baryon-photon momentum
density ratio, $R$, causes the fluid to oscillate about an offset value,
leading to a modulation in the heights of the power spectrum features with
enhanced compression (odd) and diminished rarefaction (even) peaks.
Physically, the baryons provide a ``weight'' in the baryon-photon fluid,
making it easier to fall into potential wells but harder to climb out. 

In contrast the effects of photon self-gravity cause an enhancement of the
fluctuations for $\ell\ga10^2$.
This effect is most easily understood by considering the evolution of an
overdensity in a potential well as it enters the (sound) horizon and begins
to collapse.
Since the fluid has pressure, supplied by the photons, it is hard to compress
and the infall into the potential is slower than free-fall.  Since the
overdensity is growing slowly the potential begins to decay due to the
expansion of the Universe.
The time for the photons to reach their state of maximum compression is
the same time scale for the decay of the potential, and the compressed
photons do not have a strong potential to work against as they climb back out.
Thus the  potential decay provides a near-resonant driving and
leads to a large\footnote{In the absence of neutrinos, and if the self-gravity
of the baryon-photon fluid dominates the potentials, the amplitude of the
oscillation is enhanced by $2\Psi$ on top of the $-\Psi/3$ plateau at
large scales.  In popular models the increase of a factor of 5
(from $-\Psi/3$ to $5\Psi/3$) is slightly tempered by the inclusion of
neutrinos and further diminished by the stabilizing presence of dark matter.}
increase in power \citep{HuWhi96}.

The boost from the in-phase potential decay is larger the more the self-energy
of the baryon-photon fluid contributes to the total potentials, i.e.~the
earlier in time and the smaller the (stabilizing) dark matter contribution to
the potentials.  The boost thus imprints a dependence on the matter
density\footnote{Or epoch of equality.  We shall assume that the radiation
density is known, so that equality depends primarily on $\omega_m$.}, in a
similar way to the peak in the matter power spectrum except that the
fluctuations in the CMB increase to smaller scales.
The matter density can thus be constrained by the heights of the higher order
peaks once enough peaks are seen to disentangle the confounding effects of
baryon loading and tilt of the initial spectrum.

Since much of the constraining power of the amplitude comes from measurements
of the anisotropy with $\ell\ga10^2$, the dependence of these multipoles on
$\omega_m$ at fixed $A_s$ implies that inferences of $A_s$ at fixed $C_\ell$
also depend on $\omega_m$, though in the opposite manner.

\section{Introduction to the lensing power spectrum}
\label{sec:lensCl}

We begin with a brief review of gravitational lensing of the CMB. 
For details see e.g.~the review by \citet{Lewis2006a}.
Gradients in the gravitational potential, $\Phi$, distort the trajectories
of photons traveling to us from the last scattering surface.
The deflection angles, in Born approximation, are $\mathbf{d} = \nabla\phi$,
where the lensing potential, $\phi$, is a weighted radial projection of $\Phi$.
Among other effects, these deflections alter the CMB temperature power
spectrum.  Regions that are magnified have power shifted to lower $\ell$.
Regions that are demagnified have power shifted to higher $\ell$.
The net effect of averaging over these regions is a smoothing out of the
peaks and troughs, and a softening of the exponential fall off of the
unlensed damping tail to a power law (set by the projected potential power
spectrum and the rms of the temperature gradient).
 
The key quantity for calculating the impact of lensing on the temperature
power spectrum is the angular power spectrum of the projected potential,
$C_\ell^{\phi\phi}$, which we also call the lensing power spectrum.
Taking advantage of the Limber approximation \citep{Limber}, it can be
written as a radial integral over the three dimensional gravitational
potential power spectrum $P_\Phi$
\begin{\eq}
  \ell^4 C_\ell^{\phi\phi} \simeq 4\int_0^{\chi_\star}
  \mathrm{d}\chi\ (k^4P_\Phi)\left(\frac{\ell}{\chi};a\right)
  \left[1-\frac{\chi}{\chi_\star}\right]^2 ,
\label{eqn:lensCl}
\end{\eq}
where $\chi$ is the comoving distance from the observer, $a=a(\chi)$,
a $\star$ subscript indicates the last scattering surface,
$(1-\chi/\chi_\star)^2$ is the lensing kernel, and the power
spectrum $P_\Phi$ is defined as
\begin{\eq}
  \left\langle \Phi(\mathbf{k};a)\Phi^\star(\mathbf{k'};a)\right\rangle
  =  P_\Phi(k;a)\delta^{(D)}(\mathbf{k}-\mathbf{k'}).
\end{\eq}
To calculate $P_\Phi$ we assume a power-law primordial power spectrum
$P^p_\Phi(k)$, 
\begin{\eq}
  \frac{k^3}{2\pi^2} P^p_\Phi(k) = A_s\left(\frac{k}{k_0}\right)^{n_s-1},
\end{\eq}
where $k_0$ is an arbitrary pivot point, $A_s$ and $n_s$ are the primordial
amplitude and power law index respectively.
There are two conventions for the choice of the pivot point $k_0$:
$0.002\,{\rm Mpc}^{-1}$
(constraints from WMAP\footnote{{\tt http://lambda.gsfc.nasa.gov}}) and
$0.05\,{\rm Mpc}^{-1}$ (CAMB\footnote{{\tt http://camb.info}}).
With our assumption of a power-law spectrum the two amplitudes are related by
\begin{\eq}
  A_s({\rm CAMB}) = A_s({\rm WMAP9})\times 2.5^{n_s-1}.
\end{\eq}
We adopt the CAMB convention in this paper and henceforth write
$A_s({\rm CAMB})$ as $A_s$.  For many of our formulae we shall additionally
approximate $n_s$ by $1$, thus rendering the distinction moot.

\begin{figure}
\includegraphics[scale=0.45]{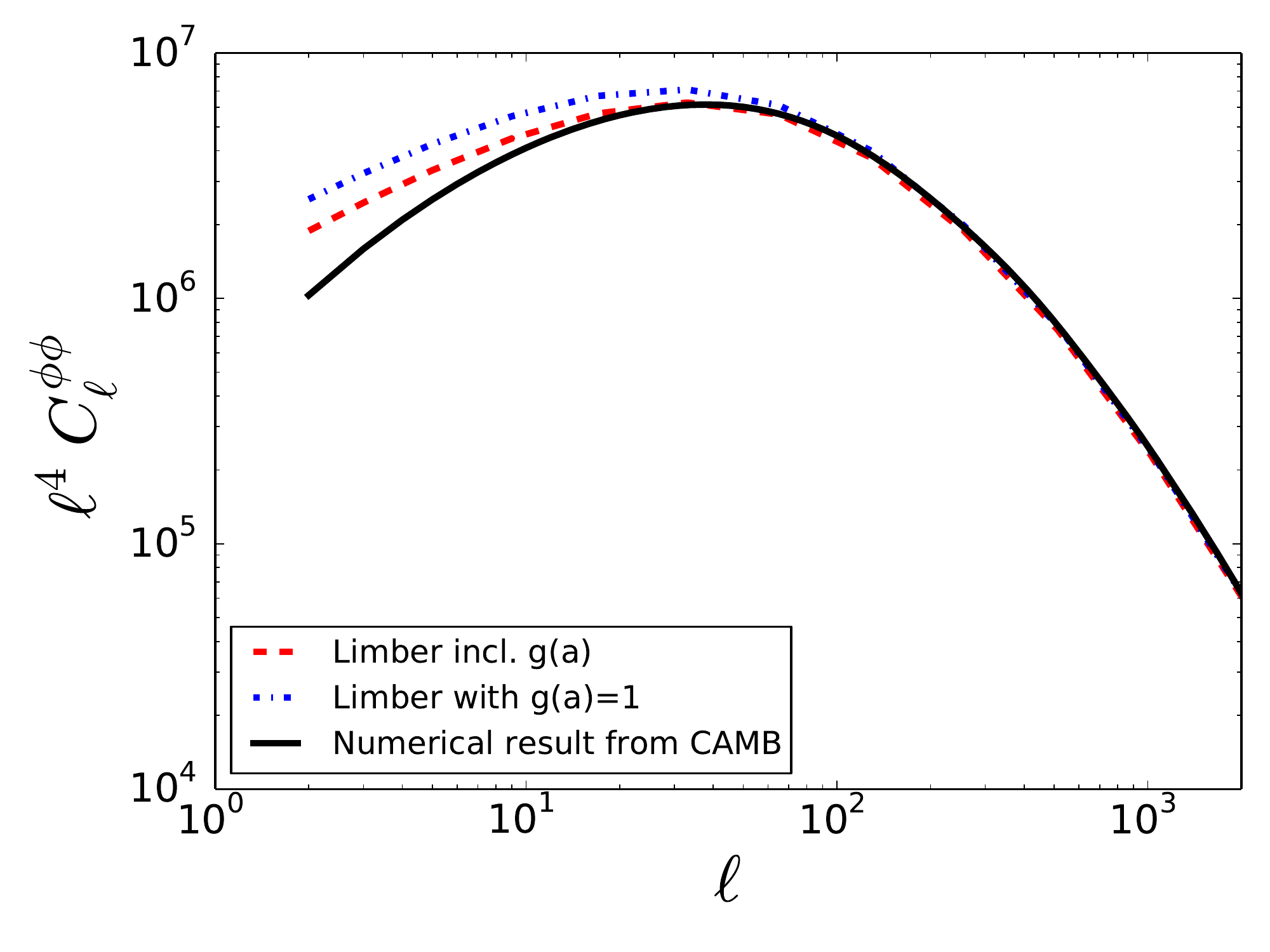}
\caption{The lensing power spectrum calculated from CAMB (solid line),
calculated with Limber approximation (dashed line) and calculated with
Limber approximation and setting $g(a)=1$ (dash-dotted line).}
\label{fig:LimberGood}
\end{figure}

The gravitational potential at late times, $\Phi(k,a)$, is related to the
primordial potential $\Phi^p(k)$ by \citep[e.g.][]{KodSas84}
\begin{\eq}
  \Phi(k,a) = \frac{9}{10}\Phi^p(k)T(k) g(a),
\end{\eq}
where the potential on very large scales is suppressed by a factor $9/10$
through the transition from radiation domination to matter domination.
For modes which enter the horizon during radiation domination when the
dominant component has significant pressure ($p\approx \rho/3$) the amplitude
of perturbations cannot grow and the expansion of the Universe forces the
potentials to decay.  For modes which enter the horizon after matter-radiation
equality (but before dark energy domination) the potentials remain constant.
The transfer function $T(k)$ takes this into account, being unity for very
large scale modes and falling approximately as $k^{-2}$ for small scales.
Once the cosmological constant starts to become important the potentials on
all scales begin to decay.  This effect is captured by the growth function,
$g(a)$, which is unity during matter domination.
With these definitions
\begin{\eq}
  (k^4P_\Phi)(k;a)=\frac{81\pi^2}{50}A_s
  \ g^2(a)\ k T^2(k)\left(\frac{k}{k_0}\right)^{n_s-1} .
\label{eqn:Pphi}
\end{\eq}

With these pieces in place, we now examine the accuracy of the Limber
approximation.
Fig.~\ref{fig:LimberGood} shows the lensing power spectrum calculated from
CAMB compared to the lensing power spectrum calculated within the Limber
approximation.  We see the agreement is very good for $\ell\ga 20$,
since the width of the kernel is much larger than the wavelength of the
modes which dominate the signal on these scales.
In order to understand the influence of the growth function $g(a)$,
we also calculated the lensing power spectrum by setting $g(a)\equiv 1$.
The growth function makes a difference only for $\ell\la 50$.
This is because for large $\ell$ the major contribution to
$C_\ell^{\phi\phi}$ comes from midway between the last scattering surface
and the observer, which is well within the matter dominated era when the
growth function $g(a)$ is close to unity. 
We are now ready to understand the impact of the matter density on
$C_\ell^{\phi\phi}$.

\section{The dependence of lensing power spectrum on matter density}
\label{sec:matter}

Eqs.~(\ref{eqn:lensCl}, \ref{eqn:Pphi}) show that there are several ways
that $\omega_m$ impacts $C_\ell^{\phi\phi}$: through the primordial amplitude,
$A_s$, the transfer function, $T(k)$, and the growth function, $g(a)$.
We will see that most of the dependence comes from the transfer function,
i.e.~from the decay of the potentials between horizon crossing and
matter-radiation equality.

\subsection{Qualitative Analysis}

Defining $x = \chi/\chi_\star$, the lensing power spectrum can be written as
\begin{\eq}
  \ell^4 C_\ell^{\phi\phi} \simeq
  A_s\chi_\star\int_0^1 \mathrm{d}x\ kT^2\left(\frac{\ell}{x\chi_\star}\right)
  (1-x)^2 g^2(a),
\label{eqn:lensClx}
\end{\eq}
where $a=a(x)$, we have dropped all constant factors and we have adopted $n_s=1$. 

\begin{figure}
\includegraphics[scale=0.45]{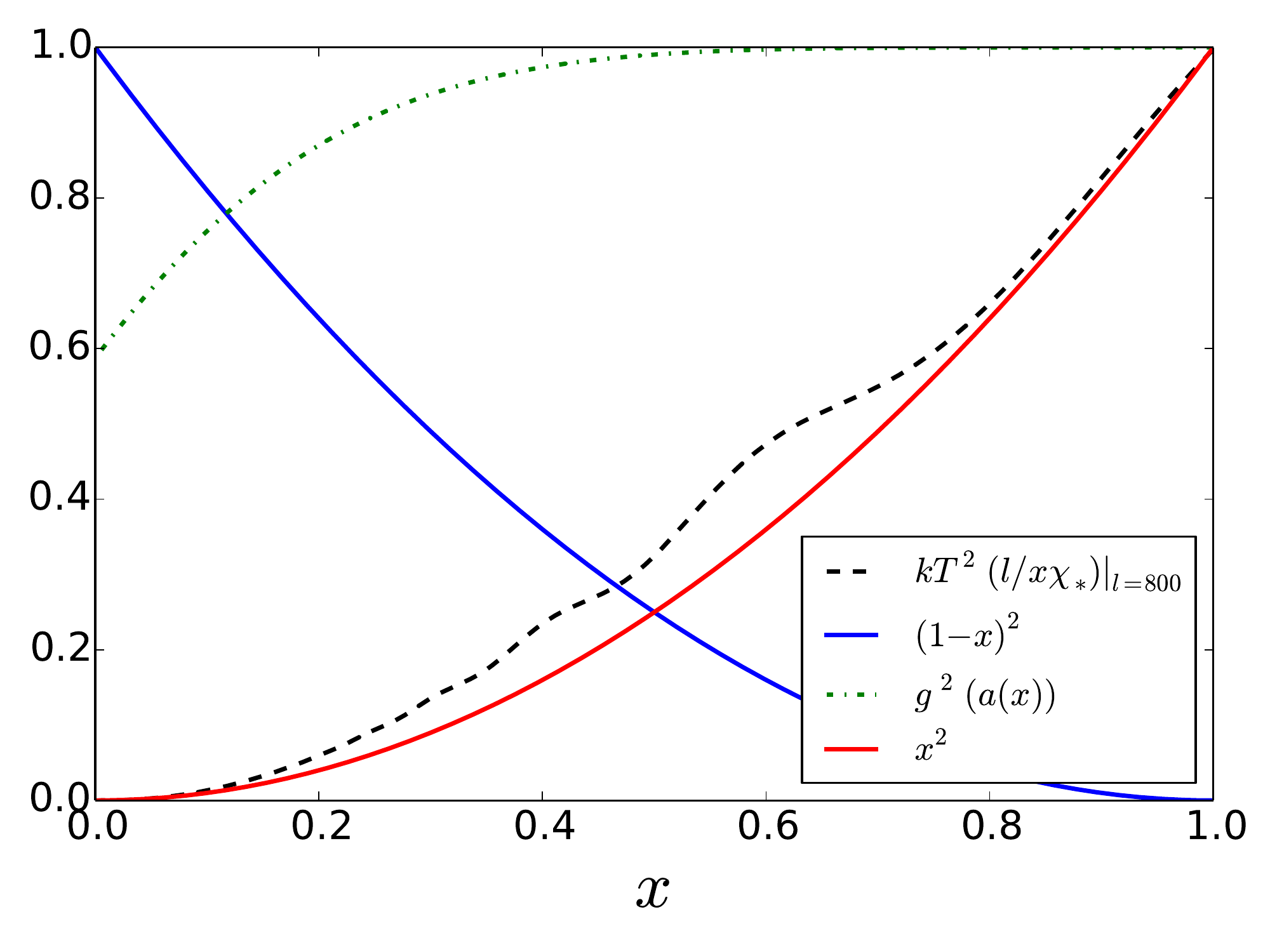}
\includegraphics[scale=0.45]{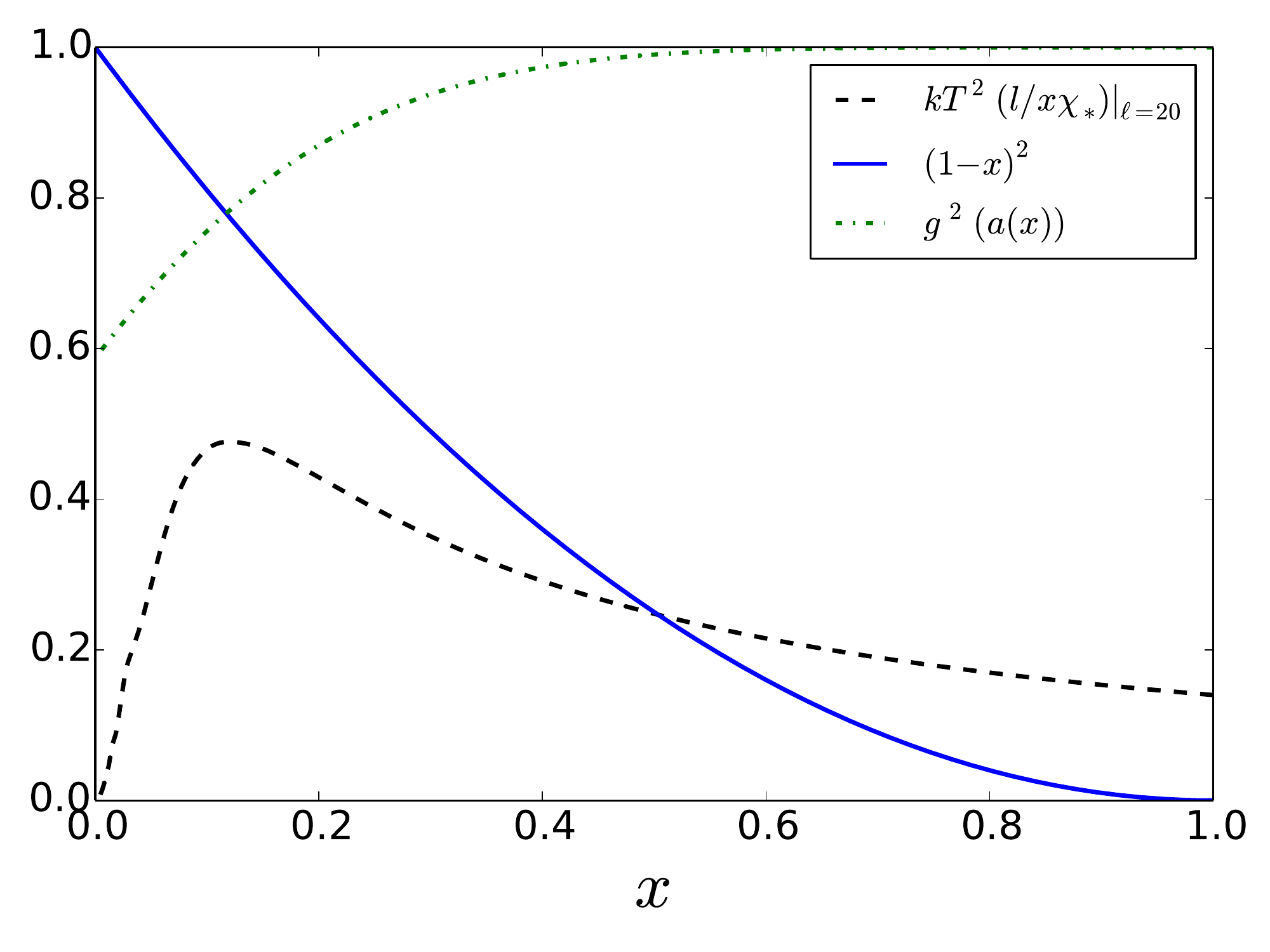}
\caption{Factors in the integrand of the lensing power spectrum for large $\ell=800$
(upper panel) and small $\ell=20$ (lower panel).}
\label{fig:integrands}
\end{figure}

We can now gain some insight by plotting in Fig.~\ref{fig:integrands}
the various factors in this integrand for two values of $\ell$,
one on each side of the $\ell^4C_\ell^{\phi \phi}$ peak ($\ell\simeq 50$).
Let's consider the $\ell = 800$ case first.
Recalling $\chi_\star\sim 10^4\,{\rm Mpc}$ we see the power spectrum
(the $kT^2$ factor, which peaks at
$k\sim k_{\rm eq}\sim 10^{-2}\,{\rm Mpc}^{-1}$)
is monotonically rising as the comoving distance, $x$, increases and we probe
power at longer wavelengths (smaller $k$).
Since the lensing kernel is dropping the net result is a broad
integrand with a peak at about 40\% of the way back to last scattering.
Very little power comes from times when the growth function is
significantly different from unity.
All dependence of the lensing kernel on $\omega_m$ has been pulled
out into the prefactor of the integrand which, as we shall see, has only
a weak dependence on $\omega_m$.
The bulk of the dependence then comes from the transfer function.  

In contrast, for the $\ell=20$ case we can see that the turnover of the matter
power spectrum (the drop in power at $k < k_{\rm eq}$) becomes very
important, reducing contributions from large comoving distance.  
In this case the growth factor and its dependence on the matter density
cannot be neglected.

\subsection{Quantitative Analysis}

To provide a quantitative test of our understanding we calculate
the dependence of $C_\ell^{\phi\phi}$ on $\omega_m$.
We begin by assuming a power-law matter power spectrum and set the growth
function to $1$.
These assumptions allow us to calculate the dependence analytically, and
they are a good approximation for
$\ell>\ell_{\rm eq}=k_{\rm eq}\chi_\star\simeq 140$.
We shall then include the physical transfer function and finally the growth
function.

We begin by determining the dependence of the prefactor in
Eq.~(\ref{eqn:lensClx}), $A_s\chi_*$, on $\omega_m$.
A sample of $\Lambda$CDM models from WMAP9 chains gives the following
scaling relations
\begin{\eq}
  A_s \propto \omega_m^{0.58}
  \quad , \quad
  \chi_\star\theta_\star \propto \omega_m^{-0.25}
\end{\eq}
where $\theta_\star$ is the angular size of the sound horizon at recombination,
which is  well determined by the WMAP9 data and is almost independent of
$\omega_m$. 
We also find the power law index $n_s$ is close to uncorrelated with
$\omega_m$, so it is safe to set $n_s=1$ as we have been doing.

The scaling of $A_s$ with $\omega_m$ comes directly from the ``potential
envelope'' effect described in Section \ref{sec:lowell}.
For the acoustic peak region the potential envelope of \citet{Hu1997}
is well fit by $\omega_m^{-0.58}$ so $A_s\propto \omega_m^{0.58}$ keeps
the power in the acoustic peaks consistent with the data.

The scaling of $\chi_\star$ with $\omega_m$ arises from the dependence of
the sound horizon on the expansion rate near last scattering.  If we assume
$z_\star$ is fixed, and if we were to neglect the contribution of radiation
to the expansion rate, then the sound horizon ($r_s$) would scale as
$\omega_m^{-1/2}$.  The presence of radiation softens this dependency to
$\omega_m^{-0.25}$ \citep{2001ApJ...549..669H}.  With $\theta_\star$ so well determined from the data,
$\chi_\star = r_s/\theta_\star$ has the same scaling as $r_s$.

The matter power spectrum $P_\delta(k)\sim kT^2(k)$ which we shall
approximate locally as a power law
\begin{\eq}
  P_\delta(k)\sim kT^2(k)
  \sim k\left(\frac{k_{\rm eq}}{k}\right)^{m+1}
  = \frac{k^{m+1}_{\rm eq}}{k^m} \quad .
\end{\eq}
The scaling with $k_{\rm eq}$ is due to the suppression of the potential
that occurs between horizon crossing and the onset of matter domination.  

\begin{figure}
\includegraphics[scale=0.45]{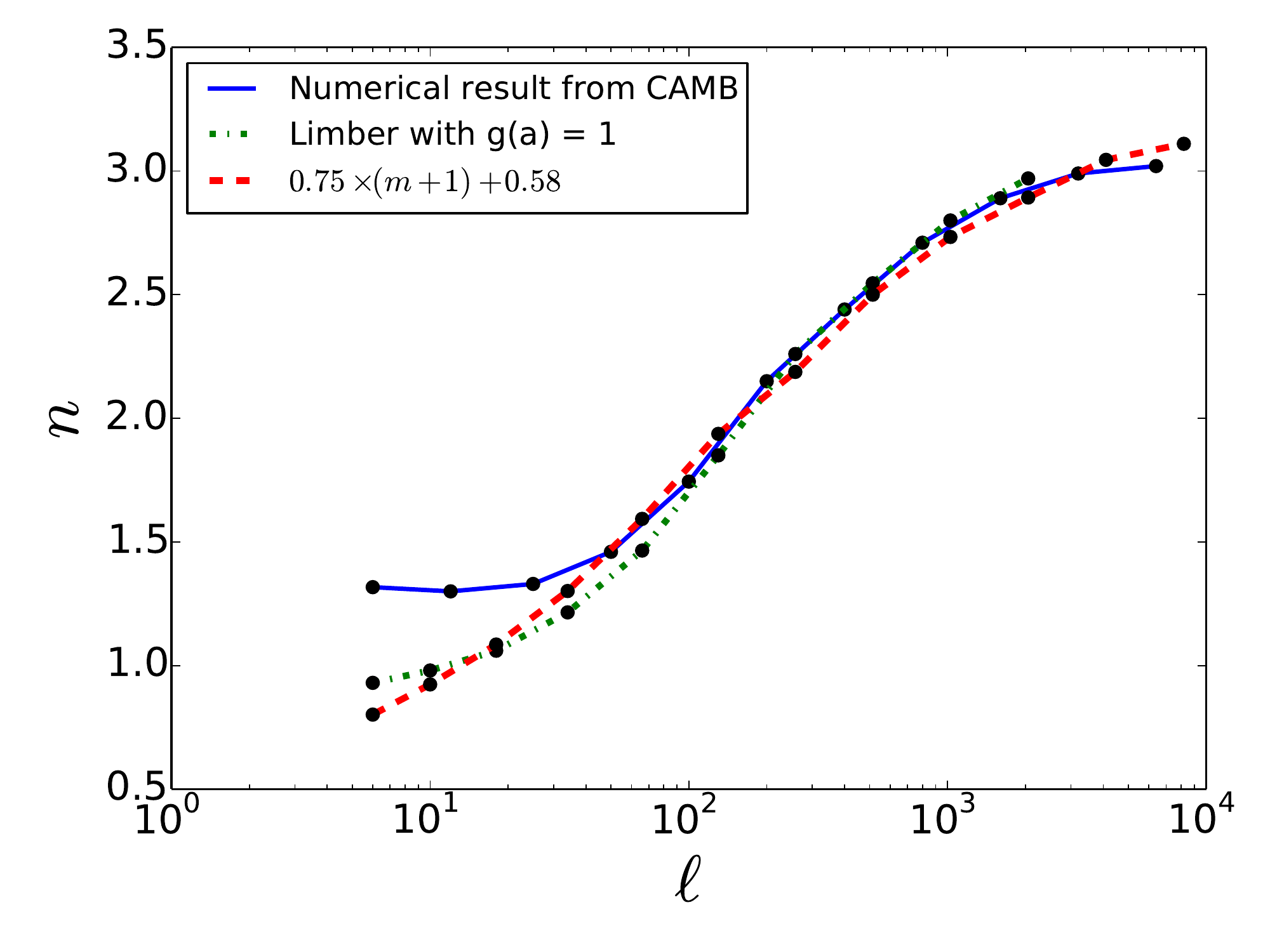}
\caption{The dependence of the lensing power spectrum on the matter
density, $\ell^4 C_\ell^{\phi\phi}\sim (\omega_m)^n$.  The solid line is the numerical result from CAMB, 
the dashed line is the result of the analytic scaling law derived in
the text ($n= 0.75(m+1)+0.58$) and the dash-dotted line is the result of
the Limber approximation setting $g(a)=1$.}
\label{fig:powerlawresult}
\end{figure}

We can now derive the dependence of $C_\ell^{\phi\phi}$ on $\omega_m$:
\begin{\eqn}
  \ell^4C_\ell^{\phi\phi}
  &\sim& A_s \frac{(k_{\rm eq}\chi_\star)^{m+1}}{\ell^m}\int_0^1
  \mathrm{d}x\ x^m(1-x)^2 g^2(a) \nn \\
  &\sim& \omega_m^{0.75(m+1)+0.58} \ell^{-m}
  \int_0^1 \mathrm{d}x\ x^m(1-x)^2 g^2(a)\nn \\
  &=& \omega_m^{0.75(m+1)+0.58} \ell^{-m}\ f(\omega_m) ,
\end{\eqn}
where we have denoted the final integral as $f(\omega_m)$ and the
$\omega_m$ dependence comes from the growth function, $g$.
Since the dependence of $g(a)$ on $\omega_m$ is strong only in the late
universe where, as we have seen, the integrand is very small the dependence
of $f(\omega_m)$ on the matter density is weak and we finally obtain the
scaling law
\begin{\eq}
  \ell^4C_\ell^{\phi\phi}\sim \omega_m^{0.75(m+1)+0.58}\ \ell^{-m}.
\end{\eq}

Of course the power spectrum is not well approximated by a single power-law.
We compute the local power-law index, $m$, numerically by matching
the numerically-determined $\ell^4 C_\ell^{\phi\phi}$ to $\ell^{-m}$.
Using this $m$ our analytic result for the scaling of $\ell^4C_\ell^{\phi\phi}$
with $\omega_m$ is very accurate for $\ell>\ell_{\rm eq}$, as we show
in Fig.~\ref{fig:powerlawresult}.

We now demonstrate that the failure of the above prediction for
$n(\ell)$ at $\ell\la 50$ is mostly due to neglecting the growth factor.
We do so by showing how the agreement with the numerical result improves
dramatically at $\ell\la 50$ if we compare to the numerical result with
$g(a)$ set to unity.
Reducing $\omega_m$ leads to an increase in $\Omega_\Lambda$ in order to
keep the angular size of the sound horizon fixed, which in turn leads to
a decrease in $g(a)$.
Thus including the growth factor increases $n(\ell)$ over the range in
$\ell$ where the growth factor is important.

\section{Discussion}
\label{sec:discussion}

\begin{figure}
\includegraphics[scale=0.45]{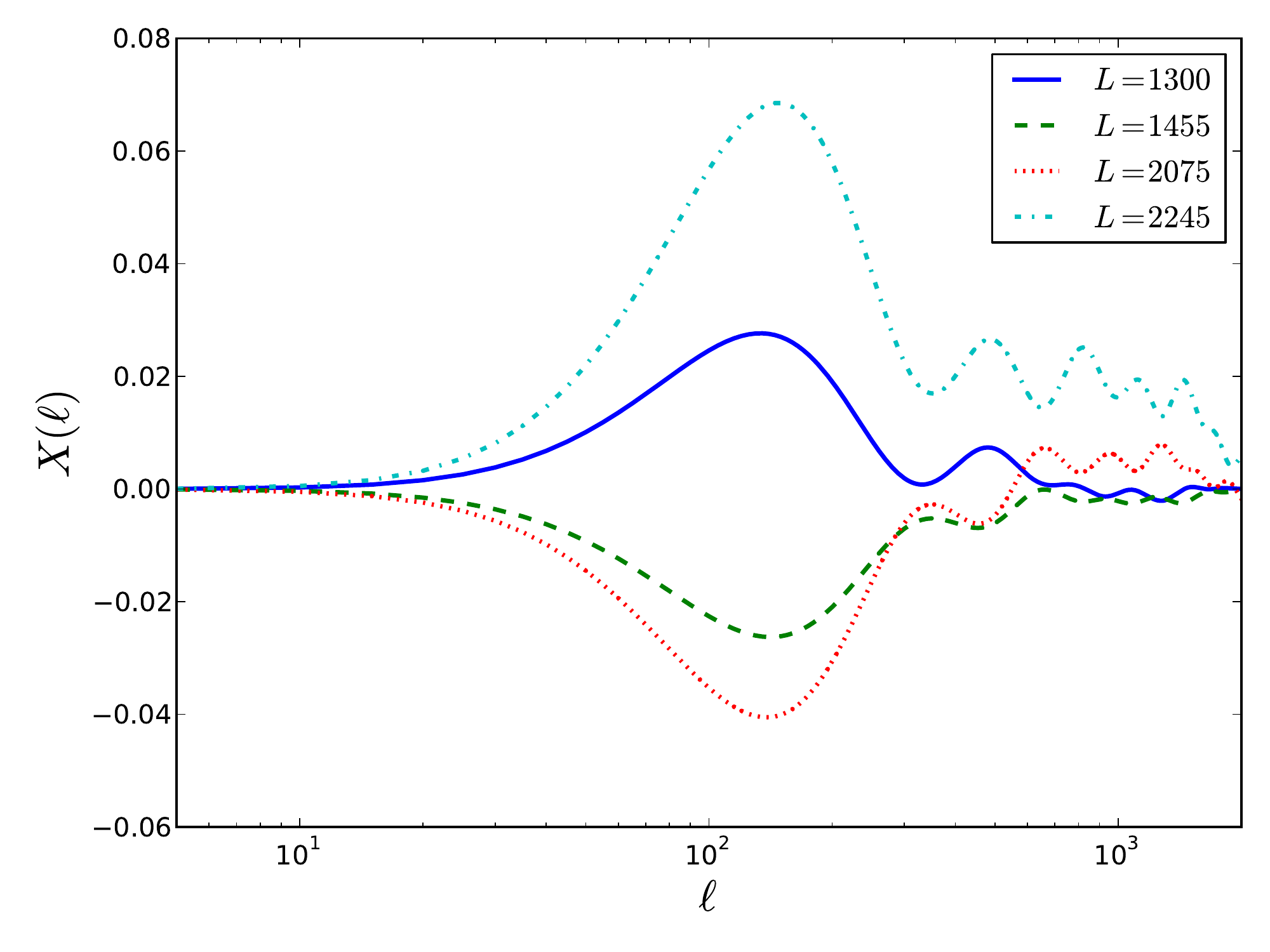}
\caption{The contribution of each multipole of the CMB lensing power spectrum, $C_\ell^{\phi \phi}$ to the
lensing of the CMB temperature power spectrum, $C_L^{TT}$, assuming our fiducial cosmology ($\rm WMAP9$ best-fit $\Lambda$CDM model), 
for two values of $L$ near acoustic peaks (1455 and 2075), and two values near troughs (1300 and 2245).}
\label{fig:weight}
\end{figure}

Precise measurements of anisotropies on small angular scales allow tight
constraints on the matter density due to its impact on the amplitude of
gravitational lensing of the angular power spectrum of temperature
anisotropies \citep{2014ApJ...782...74H}.
Within the context of $\Lambda$CDM models we found that this dependence on
the matter density arises mainly from the dependence of $P_{\Phi}(k)$ on $\omega_m$.

The origin of the dependence of $P_\Phi(k)$ on $\omega_m$ is well
understood \citep{KodSas84,Peacock99,LidLyt00,Dodelson03}, and arises from
the decay of the potential that occurs after horizon crossing but before
matter-radiation equality.
Increasing $\omega_m$ increases $z_{\rm eq}$ and thereby decreases the
amount of this decay.

Other sources of dependence (the lensing kernel, the correlation
between $A_s$ and $\omega_m$ and the growth function) contribute in a
subdominant manner.
The growth factor is only important at $\bf \ell\la 50$.
On such large angular scales the turnover of the matter power spectrum
suppresses contributions from earlier times, thereby making nearby
contributions relatively more important.

\textbf{This brings up the question, how much does $\ell < 50$ contribute to the lensing of the temperature power spectrum?
To answer it we define $X(\ell)$ via $(C_L^{TT}-C_{L \ \rm unlensed}^{TT})/C_{L\ \rm unlensed}^{TT} =  \int X(\ell) d\ln\ell$ 
(using Eq.(4.12) of \citet{Lewis2006a}) and plot the 
integrand in Fig.~\ref{fig:weight} for several values of $L$. 
For our fiducial $\Lambda$CDM cosmology, we see that $\ell < 50$ contributes a subdominant portion to these integrals. 
Setting $C_\ell^{\phi \phi}$ to zero at $\ell < 50$ 
changes the lensed $C_\ell^{TT}$ by less than $0.7\%$ at all $\ell$ with peak effect at $\ell \approx 2240$.
For comparison, setting $C_\ell^{\phi \phi}$ to zero at all $\ell$ (i.e., turning off lensing) changes the lensed
$C_\ell^{TT}$ by about $11\%$ at $\ell\approx 2240$.
Thus the lensing of the CMB temperature power spectrum has some sensitivity to the response of $g(a)$ to changes
in $\omega_m$, but this effect is subdominant to the main effect of the response of $P_\Phi(k)$ to changes in $\omega_m$.}

We note that recent measurements of $C_\ell^{TT}$ at high $\ell$ have led to small,
but important, shifts in cosmological parameters \citep{PlanckCollaborationXVI.2013}.
Assuming the minimal, six-parameter $\Lambda$CDM model, estimates of the
matter density have gone up (and estimates of $H_0$ have gone down) with
the inclusion of high $\ell$ data from Planck.
When the Planck data are restricted to the angular scales well-measured by
WMAP these shifts largely disappear \citep{PlanckCollaborationXVI.2013} suggesting
that the parameter shifts are due to the influence of the new data at small
angular scales.  As \citet{2014ApJ...782...74H} point out, the impact of
lensing on the CMB temperature power spectrum provides some of the sensitivty
to $\omega_m$ for the high $\ell$ data.
With this work we now understand the physical origin of that dependence on
$\omega_m$.  

\section{Acknowledgments}
Z.P. would like to thank Brent Follin and Marius Millea for sharing codes.
M.W. is supported by NASA.
This work made extensive use of the NASA Astrophysics Data System and
of the {\tt astro-ph} preprint archive at {\tt arXiv.org}.

\bibliographystyle{mn2e} 
\bibliography{ms}

\end{document}